\documentstyle[pra,aps,epsfig]{revtex}
\begin{document}
\language0
\tighten
\draft
\title{Photon trains and lasing : \\ The periodically pumped quantum dot}
\author{Christian Wiele\footnote{Electronic address:\
        christian.wiele@uni-essen.de} 
and Fritz Haake}
\address{Fachbereich Physik der Universit\"at-GH Essen, \
  D-45117 Essen, Germany}
\author{Carsten Rocke and Achim Wixforth}
\address{Sektion Physik der LMU,Geschwister-Scholl-Platz 1, 
   D-80539 M\"unchen, Germany}
\date{\today}
\maketitle
\begin{abstract}
We propose to pump semiconductor quantum dots with surface acoustic waves 
which deliver an alternating periodic sequence of electrons and holes. In 
combination with a good optical cavity such regular pumping could entail 
anti-bunching and sub-Poissonian photon statistics. In the bad-cavity limit a 
train of equally spaced photons would arise.
\end{abstract}
\section{Introduction}
Semiconductor quantum dots have an interesting potential for quantum optical 
applications. The growth of dots with transition frequencies in the optical 
range is very well controlled \cite{Bim96b}. Such a zero dimensional system 
leads to much higher gain than bulk or 2D quantum well structures, as shown 
theoretically as well as experimentally \cite{Bim96b,Bim96a,Bim97,Sue86}. 
Dots as active media in semiconductor lasers have already been established 
and even lasing of a single dot in a semiconductor microcavity can be achieved
\cite{Sug96,Sho96,Lott97}. From a theoretical point of view, the discrete 
states allow to treat dots much like atoms. This makes for a much simpler 
situation than, for example, the continua of states in quantum wells. 
Furthermore the semiconductor samples are small compared to atomic beams or 
even clouds of trapped atoms.\\
If a dot is to be operated as a low-noise light source it had better be pumped
in an as regular manner as possible. Yamamotos scheme of regularizing an 
injection current by a large resistor \cite{Yam87,Yam88} could hardly be 
directed to a single dot. In contrast a surface acoustic (SAW) wave could 
periodically deliver electrons and holes at a well localized array of dots or 
even a single dot \cite{Roc97}.\\
The paper is organised as follows. In section II we explain the idea
for the pumping mechanism and its theoretical implementation. In section III 
and IV we give two quantum optical applications of the system, the photon 
train and the microlaser. We end up with a conclusion and outlook in 
section V.
\section{Pump mechanism}
\begin{figure}
\begin{center}
\epsfig{file=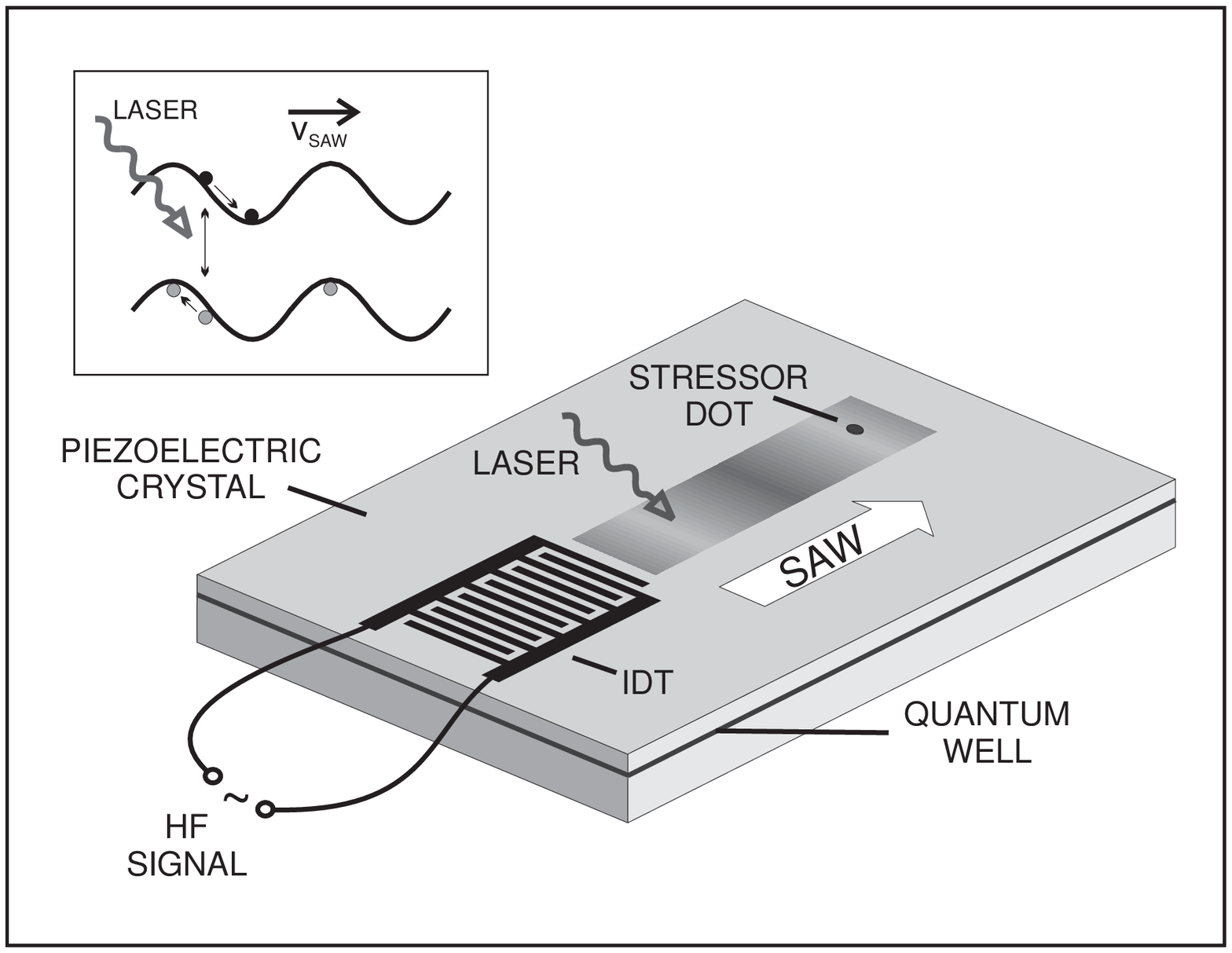,width=15truecm}
\vspace{2truecm}
\caption[FIG1]{Schematic sketch of a SAW sample. The material of the system
may be for example GaAs for the piezoelectric crystal, InGaAs for
the quantum well and InP for the stressor. The inset depict the storage
of optically generated excitons in the potential of the surface acoustic wave.}
\label{FIG1}
\end{center}
\end{figure}

To briefly explain our concept, let us consider a semiconductor quantum well 
surrounded by a piezoelectric material with an interdigital transducer (IDT) 
on top of the crystal (Fig. \ref{FIG1}). A mechanical SAW is generated by 
applying a HF signal to the IDT. The fundamental acoustic wavelength 
$\lambda_0$ and the frequency $f_0=v/\lambda_0$ are established by the 
interdigital electrode spacing, where $v$ is the sound velocity of the 
crystal. With $\lambda_0 \sim 1-3 \mu m$ and $v \sim 3 kms^{-1}$, frequencies
in the GHz range are achievable. The acoustic wave is accompanied by a 
piezoelectric field which gives an additional potential for electrons and 
holes and so periodically modulates the band edges. For high enough SAW 
amplitudes, optically generated excitons in the quantum well will be 
dissociated by the piezoelectric field (inset of Fig. \ref{FIG1}). 
A field strength of the order of $500 V/cm$ suffices and results in a wave 
amplitude of $50 - 150 meV$, depending on the wavelength. Carriers are then 
trapped in the moving lateral potential superlattice of the sound wave and 
recombination becomes impossible: Electrons will stay in the minima of the 
wave, while holes move with the maxima \cite{Roc97}. A simple estimate of 
the spatial width $\Delta d$ of the lateral ground state in the wave 
potential yields $\Delta d / \lambda < 0.02 $. We thus obtain a series of 
equally spaced quantum wires moving in the plane of the quantum well. The 
length of these wires is given by the width of the IDTs, typically 
$300 \mu m$. The occupation of the wires with electrons and holes can be 
controlled by the pump strength of the laser and is of the order $10^3 - 10^4$
carriers per wire. \\
A quantum dot for our purposes may be established by a stressor on top of 
the crystal which causes a local potential minimum in the quantum well 
underneeth. The linear dimension of typical stressor dots with transition 
frequencies in the optical range is about $10 - 30 nm$ while their potential 
depths are about $100 meV$ for electrons and $50 meV$ for holes. For further 
investigations, we assume that there is only one electron and one hole state 
in the dot. When both states are occupied an exciton is formed, so there is 
just a single exciton state. We should speak of an exited, a semi-excited, 
and an unexcited dot when an exciton, only one carrier (electron or hole), 
and no carrier is present. An excited/unexcited dot may then be treated as 
simple two-level system with pseudo-spin operators $S^+, S^-$ creating and 
annihilating an exciton. This system may interact with a single-mode light 
field. In the semi-excited case no interaction with the light field is 
possible and the creation of an exciton is only possible by capturing the 
missing carrier. \\
While being crossed by a moving quantum wire an empty dot may pluck one of
the carriers offered: If the dot potential is deep enough a carrier will drop 
into it and stay there, while the wave is moving on.\\
The scheme just sketched may indeed produce the designed properties of the 
pump. First, the periodicity of arriving carriers is given by the SAW, as the
moving wires are well separated. Second, with a density of $\approx 3$ 
carriers per $100 nm$ in a wire, there is a high probability for the dot to 
capture an electron or hole within the crossing time of a wire. Of course, a 
single dot makes but inefficient use of the moving wires, as only one of 
$10^4$ carriers is used per cycle. If one had several dots lying in a row 
parallel to the wires, better pump yields could arise. 
Another way to increase efficency may be focusing the SAW onto one or few dots,
which seems to be feasible in an experiment.\\
Our periodically pumped dot (PPD) will in practice suffer from degradation of 
complete regularity. One cause of pump fluctuations is the finite width of the
lateral SAW ground state, as mentioned above. This leads to variations in the
instant of pumping. Pump noise also results when no carrier is plucked from a
crossing wire; this may be minimised by a high electron and hole generation 
rate in the SAW, so all wires are well occupied. We neglect both types of 
noise here and consider the case of zero pump fluctuations. \\
%
We shall now discuss our pumping scheme in the framework of the 
Jaynes-Cummings model, limiting ourselves to the single PPD. 
The system is described by the exciton-field coupling constant $g$, the field
damping constant $\kappa$, the pseudo-spin operators $S^+, S^-$ and the photon
creation and annihilation operators $a^{\dag},a$. In the interaction picture 
the master equation for the density operator of dot and cavity mode is
\begin{eqnarray} \label{master1}
\dot{\rho}& =& g [a S^{+} -a^{\dag} S^{-},\rho]+\frac{\kappa}{2}
 \{ [a,\rho a^{\dag}] + [a \rho , a^{\dag}] \}\\ 
          & =:& \Lambda \rho.\nonumber
\end{eqnarray}
Due to the regularity of pumping events we cannot work with a standard pump 
term in the master equation. We rather have to solve the problem by setting 
new initial values after every pump event. The Hilbert space is defined in 
the following way. The dot may be in one of three states, the excited 
$|e\rangle$, the unexcited (ground) $|g\rangle$, or the semi-excited 
$|se\rangle$, where the special property of the semi-excited state is 
\begin{equation}
S^+|se\rangle = S^-|se\rangle = 0, 
\end{equation}
i.e. a dot in this state cannot interact. The cavity mode is expanded in the 
basis of Fock states $|n)$. We now further assume that every pump event is 
completely incoherent and destroyes all off-diagonal elements of $\rho$. With
these assumptions, the most general density operator with the condensed 
notation $|g,n\rangle:=|g\rangle|n)$ is
\begin{eqnarray}\label{rhogen}
\rho(t)=&\sum\limits_{n=0}^\infty & 
\left( C_{e,e}^n(t)|e,n\rangle \langle e,n|\right. + 
C_{g,e}^n(t)|g,n+1\rangle \langle e,n| \nonumber \\
& & {} + C_{e,g}^n(t) |e,n\rangle \langle g,n+1| +
C_{g,g}^n(t)|g,n\rangle \langle g,n| \nonumber\\
& & {} +\left. C_{se,se}^n(t)|se,n\rangle \langle se,n|\right).
\end{eqnarray}
Starting from some initial $\rho(0)$ at $t=0$, the system evolves according to
the master equation (\ref{master1}) until the first pumping event immediately
before which we have
\begin{equation}\label{freeevol}
\rho(T/2-0)=e^{\Lambda T/2} \rho(0),
\end{equation}
New initial values at $t=T/2+0$ are now set by 
\begin{eqnarray}
& &(\mbox{\bf{a}})\quad
|se,n \rangle \langle se,n|\rightarrow|e,n\rangle\langle e,n|,\nonumber \\
& & (\mbox{\bf{b}})\quad
|g,n\rangle \langle g,n|\rightarrow|se,n\rangle \langle se,n|,\nonumber\\
& &(\mbox{\bf{c}})\quad
|e,n\rangle \langle e,n| \rightarrow |e,n\rangle \langle e,n|,  
\end{eqnarray}
all other terms in (\ref{rhogen}) vanishing. The three processes indicated 
have to be interpreted like this: A semi-excited dot is excited by capturing 
the missing carrier (a), an unexcited dot captures a carrier and becomes 
semi-excited (b), and if the dot is excited at the instant of pumping, no 
pump event may occur and the system keeps the old state (c). We thus find the
new initial state after pumping 
\begin{eqnarray}\label{after}
\rho(t_i+0)=&\sum\limits_n&
\left[\left(C_{se,se}^n(t_i-0)+C_{e,e}^n(t_i-0)\right) |e,n\rangle \langle e,n|\right. \nonumber \\
& & + {} \left. C_{g,g}^n(t_i-0)|se,n\rangle \langle se,n|\right],
\end{eqnarray}
for $t_i = T/2, T, 3T/2 ...$ being the instants of pumping. Between the 
pumping events, the system evolves again like (\ref{freeevol}).
\section{Photon trains}
As a first application of this new pumping mechanism, a PPD inside a bad 
single-mode cavity is considered. We thus assume the cavity-damping rate 
$\kappa$ to be larger than the coupling constant $g$. Let us start with an 
excited dot and the light field in the vacuum state. The system will 
then undergo damped Rabi oscillations until the generated photon has left the
resonator and the dot is back in the ground state. Now we refill the dot with
an exciton (first an electron and then a hole) and the process starts again. 
By doing so, the Hilbert space of the field is confined to the vacuum $|0)$ 
and the single-photon Fock state $|1)$, i.e. there is at most one photon
in the resonator. With these assumptions equation (\ref{master1}) is exactly 
solvable. In the overdamped case $(4g<\kappa)$, where a photon is not 
re-absorbed after emission, we obtain for the probability of finding a photon 
in the resonator for a single process  
\begin{eqnarray}\label{p1}
C_{g,g}^{1}(t)&\equiv&p_1(t)\\ 
&=&\frac{8g^2}{\kappa^2-16g^2}e^{-\kappa t/2} 
      \left( \cosh (\frac{t}{2}\sqrt{\kappa^2-16g^2})-1\right).\nonumber 
\end{eqnarray}
The long-time behaviour of (\ref{p1}) is
\begin{equation}
p_1(t) \rightarrow e^{-4g^2t/\kappa}
\end{equation}
thus the pumping time $T$ has to be much larger than $\kappa/4g^2$ to ensure 
the photon having left the cavity before the next electron or hole drops 
into the dot. With these requirements met, the solution for the periodically 
excited system is
\begin{equation}
p(t)= \sum_{m=0}^{\infty} \
 p_1 (t-mT)\,\Theta\!\left( T-| 2t-(2m+1)T| \right),
\end{equation}
with $\Theta (x)=0$ for $x\leq 0$ and $1$ for $x>0$.
We call this periodic series of 1-photon processes a 'photon train', having 
the picture of a long train with equidistant waggons in mind. Note particually
that in our system the resonator serves only to enhance the coupling constant
$g$ and to orient the emission, not for accumulating photons. \\
The mean photon number $\bar{n}$ in the cavity is given by the time average of
$p(t)$ over one period $T$
\begin{equation}
\bar{n} =\frac{1}{T} \int\limits_0^Tdt \,p(t)= \frac{1}{\kappa T}.
\end{equation}
As this system is very simple most coherence and correlation properties can be
calculated analytically. For example the first-order coherence function
$g_1(\tau)=\overline{\langle a^{\dag}(t)a(t+\tau)\rangle}/
\overline{\langle a^{\dag}(t)a(t)\rangle}$ yields
\begin{equation}\label{g1}
g_1(\tau)=\sqrt{1+\kappa^2/4g^2}e^{-\kappa |\tau|/2} \cos (g|\tau| + \phi), 
\end{equation}
with $\phi = \arctan (\kappa/2g)$.
\section{PPD Microlaser}
The second case to consider is a PPD inside a high-Q single-mode cavity, so
photons may be accumulated. We do not want to present a detailed calculation
for this model here, as this system does not provide too much new. We will
rather give a comparison with a standard microlaser model for atoms
\cite{Fil86,Bergou89,Haake89,Bergou92,Puri96}.\\
In the standard model, a beam of regular distributed three-level atoms goes 
through an excitation region just before entering a single-mode cavity. 
Each atom has 
the probability $p_A$ of being excited from its ground level $c$ to the upper
level $a$. The lasing transition involves level $a$ and the intermediate
level $b$ thus an atom in level $c$ cannot interact with the light field.
Furthermore it is assumed that at most one atom is in the cavity
at a time and that the interaction time $t_{int}$ is much shorter than the 
cavity-decay time $\kappa^{-1}$ and the time $T_A$ between successive atoms
entering the resonator. This assumption allows for neglecting the field 
damping while the atom passes the cavity, leading to a simple Jaynes-Cummings
Hamiltonian during the interaction. In the interval $T_A-t_{int}$, when no 
atom is inside the cavity, pure field damping occurs. The parameter $p_A$ has
been used to describe different pumping statistics \cite{Bergou89}. The 
stationary solution exhibits sub- as well as super-Poissonian statistics, 
depending on a certain pumping parameter. Another feature of this model is the
generation of trapping states in the light field, where the photon number is 
limited to an upper boundary \cite{Puri96}. This is caused by the constant 
interaction time $t_{int}$, as for a specific photon number the atom leaves 
the resonator in the excited state (after one or several full Rabi 
oscillations), and no additional photon is emitted into the cavity.\\
Now we consider the PPD model, which is also described by the master equation
(\ref{master1}), but this time we assume $g$ to be much larger than $\kappa$. 
Spontaneous emission is again neglected. Starting with a given field, we look
at the dot just after a pump event. As described above, the dot is either in 
the excited or the semi-excited state. This is very much like in the atomic 
case, where the atom enters the resonator either in the exited level $a$ or 
in the non-interacting level $c$. From (\ref{after}) we may then define the 
probability $p_D(t_i + 0)$ for the dot to be in the excited state after the 
$i$-th pump event,
\begin{eqnarray}\label{pD}
p_D(t_i+0) &\equiv&{\rm Tr}(S^+S^-\rho)\nonumber \\
&=&\sum\limits_{n=0}^{\infty}\left(C^n_{se,se}(t_i-0)
+C^n_{e,e}(t_i-0)\right).
\end{eqnarray}
As the dot is always inside the cavity we have $t_{int} \equiv T/2$,  
corresponding to the case of an atom entering the cavity as the previous just
leaves. This circumstance requires to include field damping during the whole 
calculation, which for small damping will however not lead to essential 
changes in the results. From (\ref{pD}) we see that in contrast to the 
standard microlaser the probability $p_D$ for 'injecting' an excited dot 
depends on the state of the 'leaving' dot. This makes this system much more 
complicate to treat analytically. But it is clear that $p_D$ has to be 
constant in the stationary regime. The probability has to be determined 
self-consistently and is not an independend parameter as in the standard 
system.\\
For both models are very similar, it is not astonishing that we have 
numerically found all features of the standard microlaser (trapping, sub- and 
super-Poissonian statistics) in the PPD model. 
\section{Conclusion and outlook}
We presented the model of a new pumping mechanism for semiconductor quantum
dots and its applications in quantum optics. The combination of surface 
acoustic waves, quantum dot physics and cavities opens an interesting field 
of research inviting experimental and possibly new theoretical work. Single 
PPD's offer promise as indicated above. Collections of several PPD's close by
may be put to collective interaction with a single-mode light field. Then one
could think of a train of superradiant pulses or a superradiant laser. \\
Our simplifying assumptions (e.g. a two-level dot) have to be tested in 
experiments which will give advice for a more realistic model. \\

\acknowledgements
One of us (C.W.) would like to thank G. Bastian from the {\it PTB} for helpful 
hints. We further thank J.P. Kotthaus and K. Karrai for helpful discussions.   


\clearpage
\onecolumn
\end{document}